\documentclass[aps,prb,twocolumn,superscriptaddress]{revtex4-1}
\usepackage{amsmath}
\usepackage{graphics}
\usepackage{graphicx}
\usepackage{epsfig}
\usepackage{amssymb}
\usepackage{amsfonts}
\usepackage{color}
\usepackage{lineno}
\DeclareUnicodeCharacter{2061}{}

\usepackage{float}

\begin{document}

\title{Possible high temperature superconducting transitions in disordered graphite obtained from room temperature deintercalated KC$_8$}

\author{Samar \surname{Layek}}\affiliation{Institut N\'eel, Universit\'e Grenoble Alpes and Centre National de la Recherche Scientifique \\25 rue des Martyrs - BP 166, 38042, Grenoble cedex 9 France}

\author{Miguel \surname{Monteverde}}\affiliation{Laboratoire de Physique des Solides, CNRS-Universit\'e Paris-Sud UMR 8502, 91405 Orsay Cedex, France }

\author{Gast\'on \surname{Garbarino}}\affiliation{European Synchrotron Radiation Facility, 38048, Grenoble, France}

\author{Marie-Aude \surname{M\' easson}}\affiliation{Institut N\'eel, Universit\'e Grenoble Alpes and Centre National de la Recherche Scientifique \\25 rue des Martyrs - BP 166, 38042, Grenoble cedex 9 France}\date{\today }

\author{Andr\'e \surname{Sulpice}}\affiliation{Institut N\'eel, Universit\'e Grenoble Alpes and Centre National de la Recherche Scientifique \\25 rue des Martyrs - BP 166, 38042, Grenoble cedex 9 France}

\author{Nedjma \surname{Bendiab}}\affiliation{Institut N\'eel, Universit\'e Grenoble Alpes and Centre National de la Recherche Scientifique \\25 rue des Martyrs - BP 166, 38042, Grenoble cedex 9 France}\date{\today }

\author{Pierre \surname{Rodi\`ere}}\affiliation{Institut N\'eel, Universit\'e Grenoble Alpes and Centre National de la Recherche Scientifique \\25 rue des Martyrs - BP 166, 38042, Grenoble cedex 9 France}

\author{Romain \surname{Cazali}}\affiliation{Institut N\'eel, Universit\'e Grenoble Alpes and Centre National de la Recherche Scientifique \\25 rue des Martyrs - BP 166, 38042, Grenoble cedex 9 France}

\author{Abdellali \surname{Hadj-Azzem}}\affiliation{Institut N\'eel, Universit\'e Grenoble Alpes and Centre National de la Recherche Scientifique \\25 rue des Martyrs - BP 166, 38042, Grenoble cedex 9 France}

\author{Vivian \surname{Nassif}}\affiliation{Institut N\'eel, Universit\'e Grenoble Alpes and Centre National de la Recherche Scientifique \\25 rue des Martyrs - BP 166, 38042, Grenoble cedex 9 France}

\author{Daniel \surname{Bourgault}}\affiliation{Institut N\'eel, Universit\'e Grenoble Alpes and Centre National de la Recherche Scientifique \\25 rue des Martyrs - BP 166, 38042, Grenoble cedex 9 France}

\author{Fr\'ed\'eric \surname{Gay}}\affiliation{Institut N\'eel, Universit\'e Grenoble Alpes and Centre National de la Recherche Scientifique \\25 rue des Martyrs - BP 166, 38042, Grenoble cedex 9 France}

\author{Didier \surname{Dufeu}}\affiliation{Institut N\'eel, Universit\'e Grenoble Alpes and Centre National de la Recherche Scientifique \\25 rue des Martyrs - BP 166, 38042, Grenoble cedex 9 France}

\author{S\'ebastien \surname{Pairis}}\affiliation{Institut N\'eel, Universit\'e Grenoble Alpes and Centre National de la Recherche Scientifique \\25 rue des Martyrs - BP 166, 38042, Grenoble cedex 9 France}

\author{Jean-Louis \surname{Hodeau}}\affiliation{Institut N\'eel, Universit\'e Grenoble Alpes and Centre National de la Recherche Scientifique \\25 rue des Martyrs - BP 166, 38042, Grenoble cedex 9 France}\date{\today }

\author{Manuel  \surname{N\'u\~nez-Regueiro${^{* ,}}$}}
\email{Corresponding author. E-mail: manolo.nunez-regueiro@neel.cnrs.fr }
\affiliation{Institut N\'eel, Universit\'e Grenoble Alpes and Centre National de la Recherche Scientifique \\25 rue des Martyrs - BP 166, 38042, Grenoble cedex 9 France}

\date{\today }

\begin{abstract}
Although progress with twisted graphene nano-devices is boosting the superconductivity that is the consequence of their Moir\'e flat electronic bands, the immense choice for future development is an obstacle for their optimisation. We report here that soft-chemistry deintercalation of KC$_8$ breaks down graphite stacking generating a strong disorder that includes stacking twists and variable local doping. We obtain a bulk graphite whose individual crystallites have different  stackings with arbitrary twists and doping, scanning in the same sample a huge number of stacking configurations.  We perform magnetisation measurements on batches with different synthesis conditions.  The disorder  weakens the  huge diamagnetism of graphite, revealing several phase transitions. A  "ferromagnetic-like" magnetisation appears with Curie temperatures T$_0$$\sim$450K, that has to be subtracted from the measured magnetisation. Depending on sample synthesis, anomalies towards diamagnetic states appear at T$_c$$\sim$110K (3 samples), $\sim$240K (4 samples), $\sim$320K (2 samples). Electrical resistivity measurements yield anomalies for the  T$_c\sim$240K transition, with one sample showing a 90\% drop. We discuss the possibility that these (diamagnetic and resistitive) anomalies could be due to superconductivity. 


\end{abstract}

\maketitle

\textbf{1. INTRODUCTION}

Moir\'e patterns have been known to exist in bulk graphite for thirty years\cite{Kuwabara,Xhie}, but it is only in recent years that  they have started to be fabricated and their amazing properties studied. Chern ferromagnetic insulating states\cite{Chen} and associated superconductivity \cite{Huang} have been observed in nanodevices made of a  few layers of AA graphene twisted at a small magic angle. 

The physical reason behind these results is that  Moir\'e patterns develop extremely flat bands\cite{Mayou, MacDonald, Volovik}, that are responsible for these phenomena. Although the superconducting transition temperatures for these systems are in the range of liquid helium, the room for improvement is large, as the dispersionless energy spectrum of flat bands has a singular density of states, so that T$_c \sim \lambda$, instead of being T$_c \sim$ exp(-1/$\lambda$), where $\lambda$ is the coupling constant, i.e. T$_c$ is unbounded\cite{Volovik}. The search for higher T$_c$'s becomes the search of Moir\'e systems with even flatter bands. However, these bottom-up nano-devices are difficult to make and the number of the possible geometries and stackings tremendous.

As the Moir\'e patterns are natural elements of the stacking disorder of bulk graphite, and room temperature superconductivity has been claimed to exist in graphite\cite{Esquinazi}, we thus tried to develop a top-bottom procedure that would disturb the natural graphite ordering to introduce more Moir\'e prone  AA stacking.

\textbf{2. MATERIALS AND METHODS}

 \textbf{2.1.Synthesis}

Deintercalation of potassium at room temperature is a technique that allows obtaining metastable phases \cite{Freitas}, so we applied it to the graphite intercalated phase, KC$_8$, that has an AA structure separated by potassium atoms. 
Our samples were made by an intercalation-deintercalation process (IDI) on bought Highly Oriented Pyrolithic Graphite (HOPG) plates. First stage potassium intercalated graphite was produced using know techniques\cite{Herold} at Institut N\'eel at Grenoble (IN). Namely, potassium and highly oriented pyrolytic graphite were sealed in a quartz tube in stoichiometric proportions and maintained at 250$^\circ$C for four days.
The potassium was deintercalated as in Ref. [9] by immersion for ten hours with a mechanical agitator in a solution in acetonytrile of iodine, whose weight was stoichiometric with the potassium of the KC$_8$ sample. Samples were  dried, and some also pumped to clean them totally from the acetonytrile. All manipulations are done in a glove box.
Samples A and B were obtained using an IDI  on a ZYA HOPG, while for sample C (not pumped), a ZYB quality HOPG was used.
For X-rays measurements it was necessary to plunge the sample $ \textbf{BX}$ two days in water at 80$^{\circ}$C to dissolve most of the potassium iodide left in the sample, as it has strong reflections at angles similar to those of the graphite.
While KC$_8$ is golden and pyrophore, the obtained samples look like HOPG and, although more brittle and very cleavable, can be manipulated normally. Many present time evolution of their properties. Scanning electron microscope analysis, performed at IN using a (FESEM) ZEISS Ultra+, show the presence of residual potassium and iodine from the IDI process.

\textbf{2.2. Methods}

\begin{figure*}[hbt]

\includegraphics[width=\linewidth]{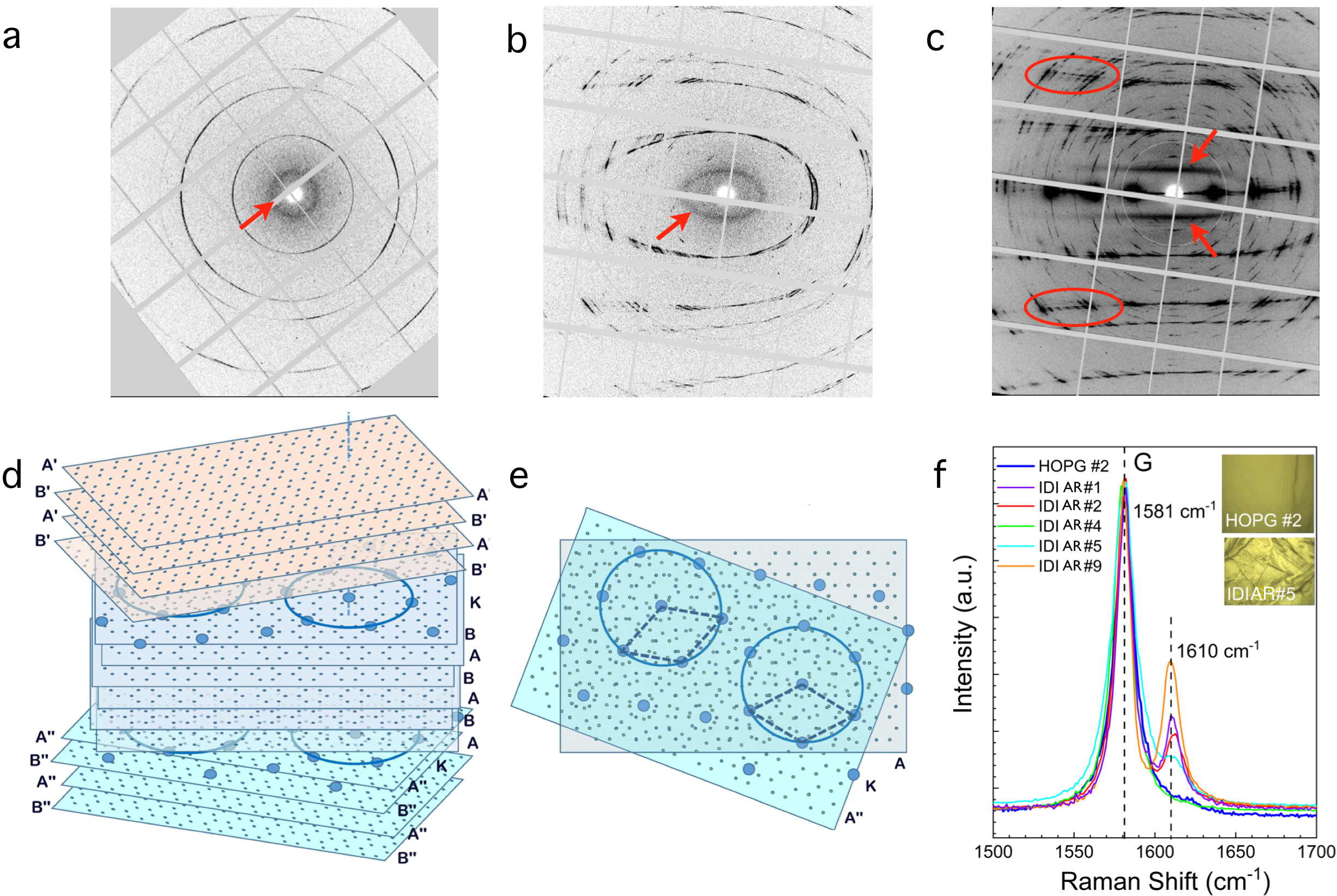}
\caption{ 
(a) X-ray pattern of IDI sample $ \textbf{BX}$ (see Methods) corresponding to an orientation close to ($\textbf{ab}$) planes ($\omega \sim$ 0-10$^\circ$) presenting, as for HOPG, a full rotation of ($\textbf{ab}$) planes  with $\textbf{a}$=$\textbf{b}$=2.46 {\AA}  parameter.  The additional diffuse ring (red arrow) is due to K short range order (SRO) of about $\sim$6{\AA}.~(b) Pattern of the inclined platelet ($\omega\sim$ 50-55$^\circ$)
 showing a deformed “ellipse”(red arrow), a section of a diffuse cylinder from the SRO.
(c) Pattern along the $\textbf{c}$ direction ($\omega\sim$ 80-90$^\circ$), 
 presenting additional modulations (red ellipses) along $\textbf{c}$, parameter 22.15 ${\AA},$ and two broad diffuse lines (red arrows) parallel to the $\textbf{c}$ direction due to the SRO diffuse cylinder of K atoms implying a total disorder of their stacking along $\textbf{c}$. (d) Example of stacking of graphite AB planes proposed for the 6 stage K deintercalated  phase; the diffuse central cylinder implies a  SRO distribution of K atoms in the ($\textbf{ab}$) plane, in-between A A planes, a total disorder along the $\textbf{c}$ direction and random rotations in the ($\textbf{ab}$) plane. (e) Example of such random rotations of A K A’ planes.
 (f) Raman spectra of the G band of one precursor HOPG and IDI sample $\textbf{AR}$ measured on different spots. The mode at $\sim$1610~cm$^{-1}$ that appears after the preparation is identified as intralayer vibrations of type B layer \cite{Solin} due to the presence of remaining K intercalation estimated at KC$_{168}$ at the surface of the sample..}
\label{Crys} 
\end{figure*}

X-rays measurements were performed on ID15b high pressure beamline at the European Synchrotron Radiation Facility (ESRF), using small X-ray beam and high energy $\lambda$=0.41\AA~and an Eiger high sensible large area detector, having a very good peak/background ratio, allowing collection of the full powder diffraction data including their diffuse contribution. We performed  measurements at each degree tilting gradually the samples from -40$^\circ$ to+89$^\circ$ with respect to the \textbf{c} axis, with 1sec acquisition per degree. The instrumental parameters (distance sample to detector, detector tilt) have been calibrated using Si powder and a vanadinite single crystal. Berylium lenses were used to focus the X ray beam to a spot size of 5$\mu$m by 5$\mu$m at the sample position. A one axis goniometer allowing to rotate the sample from -40 to +89 degrees is used to collect X ray diffraction patterns of the samples with slicing of 1 degree and 1 second acquisition time per degree. A CdTe 9M EigerII detector from Dectris has been used to collect 2D diffraction patterns and integrated using PyFai code implemented in Dioptas.

MicroRaman scattering was performed at IN at room temperature with a single stage Witec spectrometer. 
The spot size of 1~$\mu$m allows the spatial mapping of the Raman response. A solid state laser with a 532~nm line was used.

Magnetisation measurements were performed in a in a Metronique Squid Magnetometer and a Quantum Design MPMS3,VSM-Squid Magnetometer at IN, and two Quantum Design MPMS XL QD at IN and the Laboratoire de Physique des Solides (LPS). However, the best data were obtained on the Quantum Design MPMS3,VSM-Squid Magnetometer, which, by being quicker, yields more detailed data. Addenda (sample holder, straw or plastic film) were carefully measured to ensure they were not at the origin of the anomalies, and their contribution subtracted when not negligible. De Haas-van Alphen measurements were performed at LPS.

Transport measurements were performed in a home-made apparatus at IN, using either a four wire AC LR400 resistance bridge or a four wire square wave AC MM3 device, and a four wire Quantum Design PPMS at LPS, where the magnetic field and specific heat measurements were done.

\textbf{3. RESULTS}

\textbf{3.1. Structural Properties}

\textit{\textbf{3.1.1. Synchrotron X-rays measurements}}

Pieces of IDI samples $ \textbf{A}$ and $ \textbf{B}$ ($ \textbf{BX}$), and untreated HOPG used to make sample $ \textbf{B}$ were measured by X-ray diffraction at the ESRF.  From the comparison of the HOPG patterns and the IDI patterns , corresponding to nearly the same crystallographic orientations (Fig.~5abc,a'b'c' in Appendix A), it is clear that the IDI sample, although more disordered, conserves the global HOPG morphology.
Moreover, it has a 6-fold modulation K ABABAB K BABABA along $\textbf{c}$ (6 stage) with a periodicity close to 2x22.15{\AA} (red ellipses on Fig.~\ref{Crys}c). Depending to the analysed sample area, it can also be locally described by a 5-fold modulation K ABABA K ABABA along the $\textbf{c}$ axis (5 stage) with a periodicity of 19.0{\AA}. These correspond to  intercalated KC$_{72}$ and KC$_{60}$  allotropes \cite{Schulze}~(Fig.~\ref{Crys}c). The diffuse central cylinder (red arrows Fig.~\ref{Crys}a,b,c) implies a short range order (SRO) potassium distribution in the ($\textbf{ab}$) plane).  With this SRO, the K-K distances around $\sim$6\AA~in the ($\textbf{ab}$) plane, as in the liquid state\cite{DiCenzo,Clarke,Moret} found in graphite intercalates for stages L$\geq$2. This potassium long range disorder implies random doping of different regions of the sample. Our X-rays analysis suggests that the AKA stacking allows a rotation of carbon planes, determined by the variable local K occupancy, through the nearly superposition of carbon hexagons(Fig.~\ref{Crys}d)). More details can be found in Appendix A.

\textit{\textbf{3.1.2. Raman Measurements}}

MicroRaman scattering was performed at IN at room temperature (Fig.~\ref{Crys}e). Pieces of IDI samples $ \textbf{A}$($ \textbf{AR}$) and $ \textbf{B}$ , and untreated HOPG used to make sample $ \textbf{B}$ were measured on several spots. The G mode at 1580~cm$^{-1}$ is splitted after the IDI process. A second mode, G$^+$, appears at $\sim$1610~cm$^{-1}$. Its energy and intensity is dependent on the position of the laser spot, pointing to sample inhomogeneity. This G$^+$ mode is identified as intralayer vibrations of type B layer (2 stage) due to the presence of remaining K intercalation and has been quantified\cite{Solin} as a function of stages $n$. The ratio of the two modes’ intensities ($R=\frac{I_G^-}{I_G^+}$) as a function of $n$ yields the maximum remaining potassium content in the structure. Since $R$ ranges above 1.8 up to purely Carbon compound ($R=\infty$), we extrapolate\cite{Solin} a distribution from pure graphite up to a maximum K stoichiometry of KC$_{168}$ from one spot to the other. From our XRD results, each measured spot (1$\mu m^2$ surface, penetration depth of light of 50~nm) most probably encompasses a distribution of graphite and KC$_{2n}$ with $n$=2 up to 6. Consistently, the position of the G$^+$ mode decreases from 1612 to 1610~cm$^{-1}$ with decreasing ratio $R$\cite{Solin}. The ratio between the intensities of D and G peaks is inversely proportional to the graphite crystallite size L$_a$ \cite{Schwan1996}. We extract a crystallite size of 90~nm and a cluster size lower than 10\AA~  in pure HOPG and IDI samples, respectively, changing from nanocrystalline graphite to amorphous carbon \cite{Ferrari2000}.
The 2D band is enlarged in IDI samples. The smaller the $R$ ratio, the larger is the 2D band (Fig.~B1, Appendix B).

\begin{figure*}[ht]
\includegraphics[width=\linewidth]{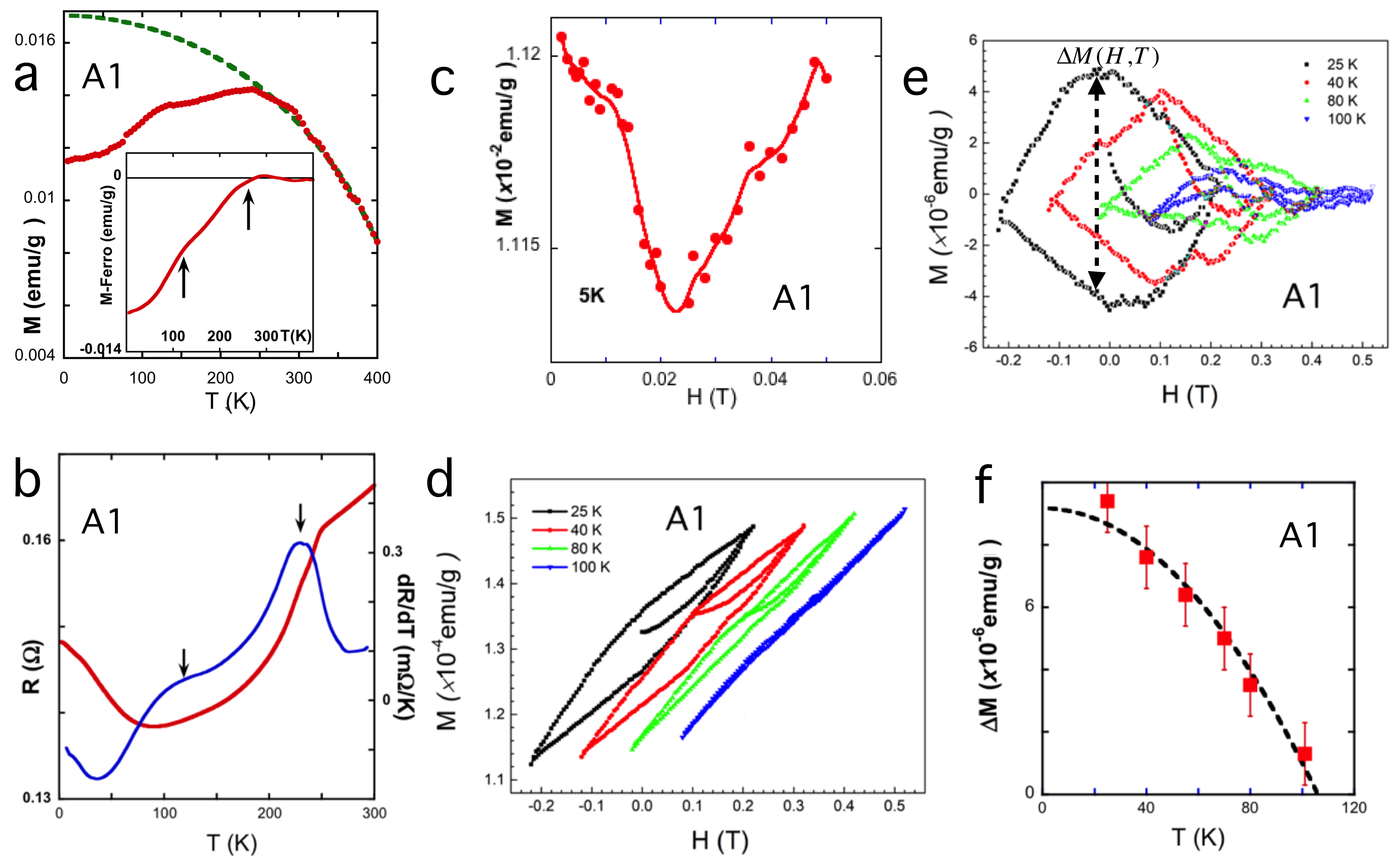}
\caption{ (a) Temperature dependence of zero field cooled (ZFC) magnetisation of sample A1(H//\textbf{c}; H=0.1T).  The green dashed line is the ferromagnetic fit using eq. (\ref{Ferro}). Inset: ZFC magnetisation as a function of temperature after subtraction of the fitted ferromagnetic contribution, showing the two diamagnetic transitions. (b)Temperature dependence of the electrical resistance and its derivative, with respect to temperature, of sample A1. (c) Magnetisation as a function of magnetic field at T=5K, showing the diamagnetic contribution.  (d) Magnetisation hysteresis cycles at different temperatures (curves are shifted by 0.1T for clarity). (e) Diamagnetic hysteresis cycles , obtained by subtracting the linear paramagnetic contribution at each temperature (curves are shifted by 0.1T for clarity). We observe that their size diminishes with increasing temperature. (f) Evolution of $\Delta$M(H,T), as defined in (d), plotted as a function of temperature for a fixed $H$=0T and fitted by expression (\ref{Jc}), showing a T$_c\sim$110K. } 
\label{R-M} 
\end{figure*}

\textbf{3.2. Physical Properties}

The deintercalation at room temperature of AA KC$_8$ has produced a highly disordered high stage intercalate with mainly a KC$_{72}$ structure, with more orientational disorder that the starting HOPG. This disorder, added to the smaller than necessary K concentration, implies that the six AB slabs have facing AA separations twisted with a range of different angles. The ideal  AKA stacking  will favour  low "magic angles" that are just a slight departure from it. Also the K concentration, smaller than that corresponding for the 6 stage ordered compound concentration, indicates that the local doping is also largely variable. Thus IDI samples presumably scan angles and dopings. Although all the synthesised samples have ferromagnetic and diamagnetic states, the portions that we measured had variable contents of different phases and we present only the data of the samples where the described state was stronger. We now show measurements that evidence three recurrent  diamagnetic transitions at T$_c$ $\sim$110K, $\sim$240K and $\sim$320K, while the ferromagnetic estimated transitions are all around T$_0\sim$450K at low magnetic field.

\textit{\textbf{3.2.1. 110K Transition.}}

 The magnetisation as a function of temperature of sample $ \textbf{A1}$ (portion of sample $ \textbf{A}$) is shown on Fig.~\ref{R-M}a. Contrary to HOPG that has strong diamagnetism, our sample follows a ferromagnetic behaviour at high temperatures. It has neat kinks at 240K and 110K(Fig.~\ref{R-M}a). 

The resistance has a linear temperature dependence at high temperatures with a sharp decrease at around $\sim$240K (Fig.~\ref{R-M}b). The decrease finishes in a power-law-like increase at low temperatures. 
 The derivative of the resistance (blue curve)
 shows two peaks at the same temperatures as the kinks in the magnetisation curve (Fig.~\ref{R-M}b).

 The ferromagnetic dependence of the magnetisation at temperatures higher than 240K can be fitted by the phenomenological function describing the temperature dependence of ferromagnetic ordering
 \begin{equation}
 M_s(T,H)=M_0(H)[1-(T/T_0)^2]^{1/2}  
 \label{Ferro}
 \end{equation} 
used previously \cite{Carley} for ferromagnetic dependences (green dashed curve on Fig.~\ref{R-M}a). $T_0$=450K is the Curie ferromagnetic temperature and $M_0$=0.017emu/gr, the full magnetisation at saturation that is equivalent to 4 10$^{-5}$ $\mu_B$ per carbon atom.

To emphasise the two diamagnetic transitions, we subtract this ferromagnetic dependence from our measurements (inset Fig.~\ref{R-M}a). It is clear from these results that the magnetisation of sample $ \textbf{A1}$ can be explained by a ferromagnetic state with $T_0$=450K and two diamagnetic transitions at T$_c\sim$110K and T$_c \sim$240K. The latter accompanied by resistivity anomalies as marked by arrows on Fig.~\ref{R-M}.

 \begin{figure*}[ht]

\includegraphics[width=\linewidth]{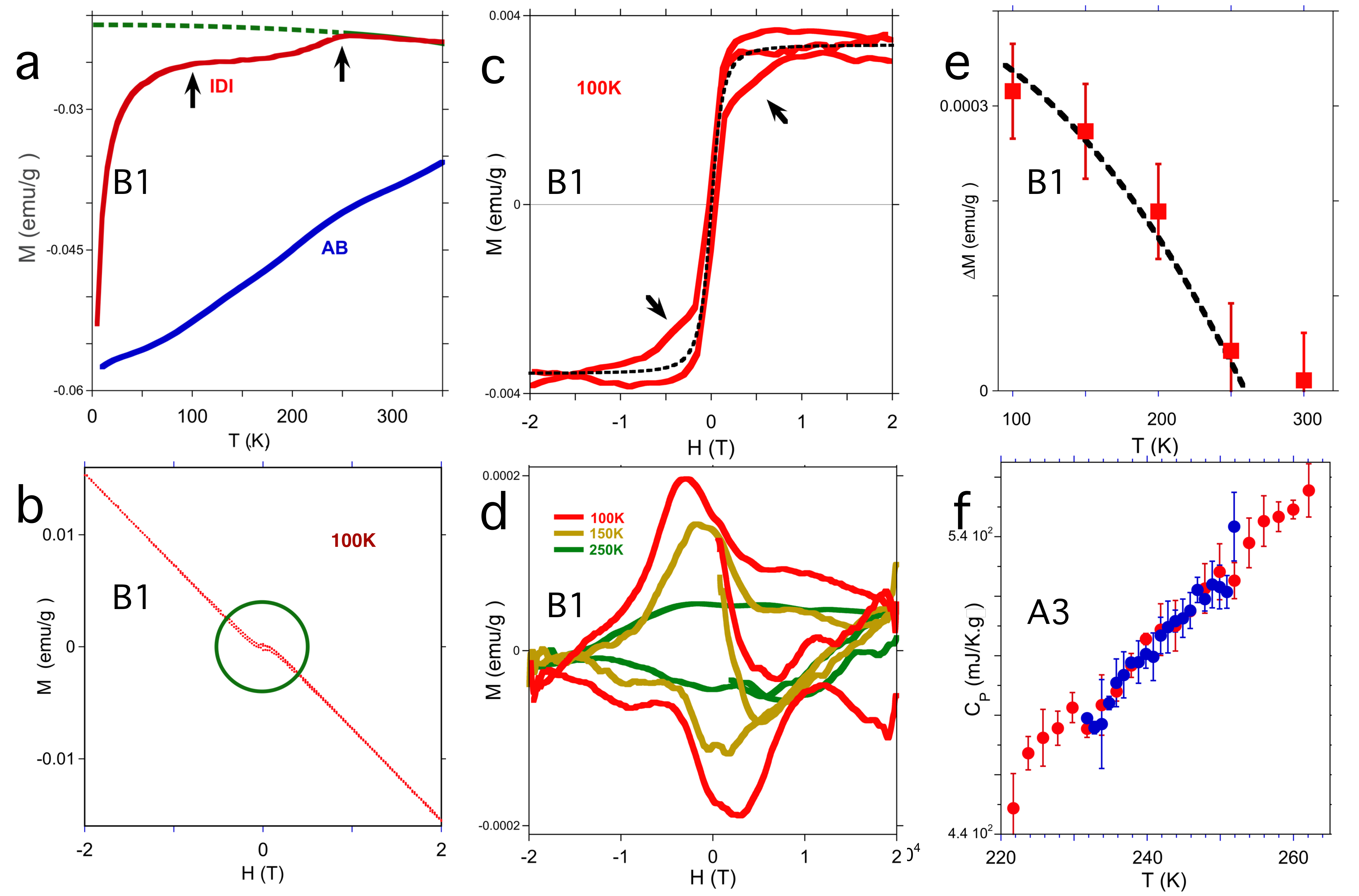}
\caption{ a) Comparison of the magnetisation (H//\textbf{c}; H=3T) as a function of temperature of virgin (AB blue curve) and intercalated and deintercalated HOPG (red curve IDI sample $\textbf{B1}$). The green dashed line is the ferromagnetic fit of the data above 245K using eq. \ref{Ferro}. (b) Magnetisation as a function of magnetic field at 100K, the green circle shows the small ferromagnetic cycle. (c) ferromagnetic cycle obtained at 100K from the magnetisation as a function of magnetic field by subtracting the diamagnetic contribution and fitted by an algebraic sigmoid function (black dashed curve). (d) Diamagnetic cycle at 100K obtained by subtracting the sigmoidal fit from the ferromagnetic cycle. (e) Evolution of  $\Delta$M at $H$=0.5T of the different cycles with temperature,  fitted by expression (\ref{Jc}), 
showing a T$_c$$\sim$245K. (f) Temperature dependence of the specific heat of sample $\textbf{A3}$, red and blue circles show different runs. We see no anomaly that would correspond to a structural transition like a CDW or other structural transition capable of explaining the behaviour of the electrical resistivity. }
\label{HyCy240} 
\end{figure*}

The ZFC moment versus magnetic field at 5K shows a diamagnetic decrease followed by a paramagnetic increase (Fig.~\ref{R-M}c). This behaviour is characteristic of superconductivity.

We have performed magnetic moment hysteresis cycles at different temperatures below 100K on sample \textbf{A1}. The cycles are sitting on a linear positive slope (Fig.~\ref{R-M}d) that we fit and subtract. We show on Fig.~\ref{R-M}e the diamagnetic hysteresis cycles for different temperatures. The cycles, that  are superconductor-like, clearly shrink with increasing temperature (Fig.~\ref{R-M}f). Superconducting cycles shrink with increasing temperature becoming totally flat at  T$_c$. To determine T$_c$ we fit the data using Bean's model\cite{Bean}. According to this model,  $\Delta$M(H,T) is proportional to the critical current $J_c$(H,T), that follows with temperature the expression (\ref{Jc}).
\begin{equation} \label{Jc}
 {J_c(H,T)=J_{c0}(H)[1-(T/T_c)^ 2]}
 \end{equation}
The fit of $\Delta$M(H,T) points to a T$_c\sim$110K, that corresponds to the low temperature transition (marked with an arrow) on Figures~\ref{R-M}a and ~\ref{R-M}b.

\textit{\textbf{3.2.2. 240K Transition.}} 

We show on Fig.~\ref{HyCy240}a the comparison of the magnetisation of the HOPG sample before and after the IDI process (sample $\textbf{B1}$). Its diamagnetism is three times smaller than that of the sample in the Bernal structure. Anomalies at $\sim$110K and $\sim$240K are once more present (arrows on Fig.~\ref{HyCy240}a). The behaviour above 240K has ferromagnetic temperature dependence and can be fitted by expression (\ref{Ferro}), with T$_0$=550K for 3T and a $M_0$ corresponding to $\sim$8x10$^{-6} \mu_B$ per carbon atom.

The magnetisation of the sample as a function of field shows a negative diamagnetic slope, with a small ferromagnetic sigmoidal jump around zero field (Fig.~\ref{HyCy240}b). After subtracting the diamagnetic contribution, the ferromagnetic cycle is seen on Fig.~\ref{HyCy240}c. 

What attracts the eye are the unusual "wings" below and above the jump (marked by arrows on Fig.~\ref{HyCy240}c).  This type of cycle shape is typical of mixed ferromagnetic-superconducting materials\cite{Scheidt}. As is usual in these cases\cite{Ursula}, an algebraic sigmoidal function was fitted and subtracted from the data.  The resulting curves are diamagnetic cycles (Fig.~\ref{HyCy240}d). The temperature evolution of $\Delta$M(T) is plotted  on Fig.~\ref{HyCy240}e.  $\Delta$M(T) decreases with increasing temperatures, pointing towards a transition at around 250$\pm$10K.Within this error, our analysis points towards a T$_c\sim$ 240K also seen on the magnetisation of Fig.~\ref{HyCy240}a.

\textit{3.2.2.1. Effect of magnetic field on the transition}

\begin{figure*}[ht]
\includegraphics[width=\linewidth]{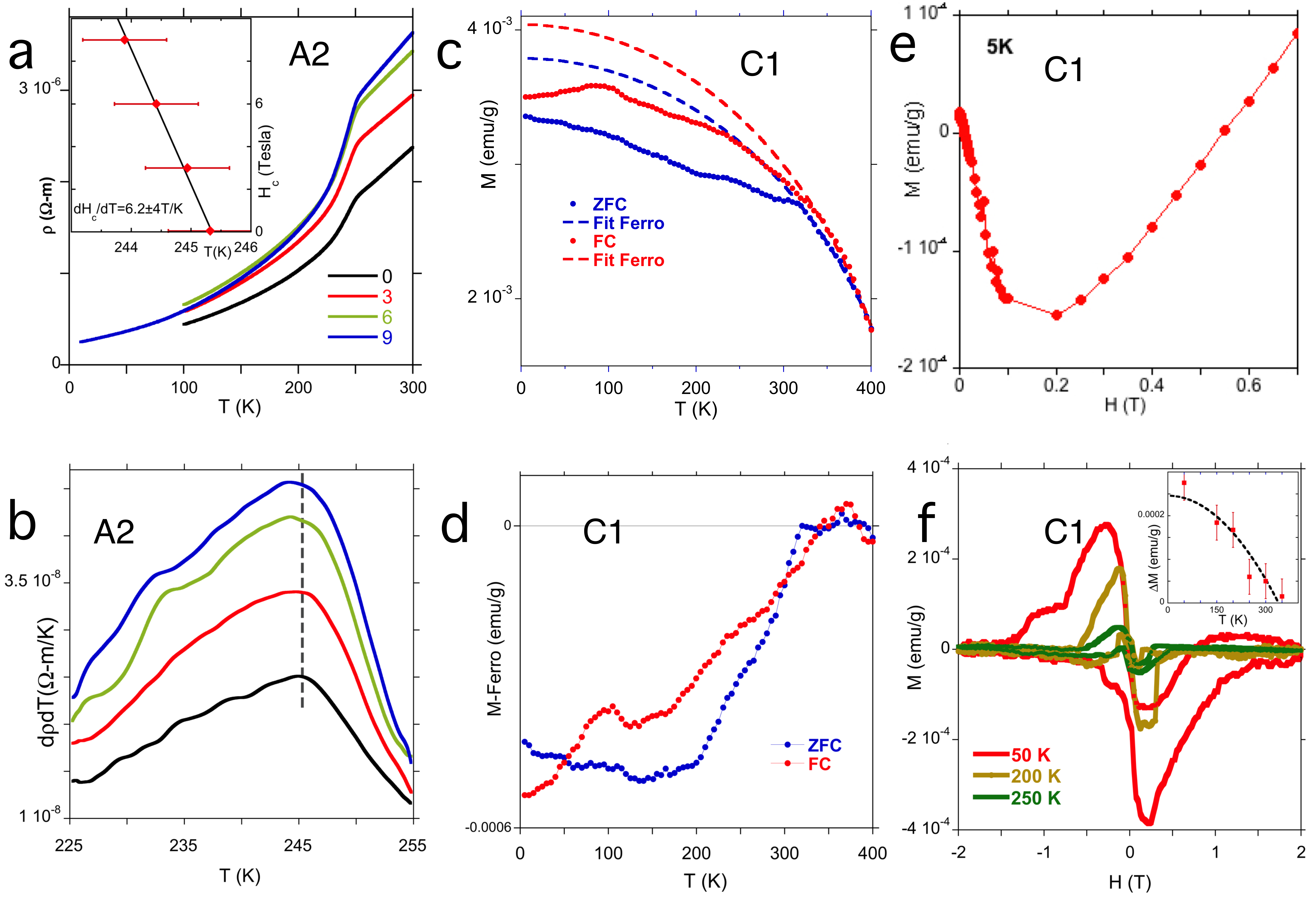}
\caption{ (a)  Temperature dependence of the electrical resistivity of sample $\textbf{A2}$ as a function of magnetic field. (b) Derivative of the resistivity as a function of temperature for different fields showing that up to 9T. The vertical dashed line aids to visualise the very small shift of the maximum with field. (c) ZFC and FC curves(H//\textbf{c}; H=2T) of sample $\textbf{C1}$ at a magnetic field of 2T. Above $\sim$ 320K they both follow a ferromagnetic behaviour shown by the dashed curves. (d) the same curves after subtraction of the ferromagnetic behaviour. (e) Magnetisation curve as a function of magnetic field after annealing at 400K. (f) Diamagnetic hysteresis cycles at 50, 200 and 250K after subtraction of the bulk paramagnetic behaviour; (inset) $\Delta$ M(H,T) plotted as a function of temperature for $H$=0.2T,  fitted by expression (\ref{Jc}), 
showing that the transition temperature is T$_c \sim$320K.
 }
\label{ZYB} 
\end{figure*}

We also measured the magnetic field effect on the transition at T$_c$ $\sim$240K. We used  the electrical resistivity of sample $\textbf{A2}$ (Fig.~\ref{ZYB}a), that has a rate of percolation of more than 90\% for this transition. There is a strong positive magnetoresistance of the background forcing the use of the derivative with respect to temperature of the electrical resistivity to verify the field effect on T$_c$. 
The derivative shows a peak, corresponding to T$_c$, that decreases with magnetic field by less than 2K for 9T, as plotted in the inset of Fig.~\ref{ZYB}a. The dependence is very steep for the few points that we have, although we can extract a slope with a very large error dH$_c$(T)/dT =6.2$\pm$2T/K. In the hypothesis that the resistivity jump is due to superconductivity, using the Werthamer-Helfand-Hohenberg formula we obtain for the critical field  at zero temperature a value H$_c$(0)=-0.69 T$_c$dH$_c$(T)/dT=1048$\pm$200T. 
This huge value scales with what has been measured on Moir\'e multilayers\cite{Watanabe,Park}, i.e. H$_c$(0)$\sim$4-5Tesla per Kelvin of T$_c$, instead of $\sim$1 for most superconductors, suggesting common origin . The coherence length is estimated to be $\xi$(0) =\{$\Phi_0$/[2$\pi$H$_{c}$(0)]\}$^{1/2}$=0.5$\pm$0.1nm. These results are preliminary as much higher fields are necessary to correctly quantify the effect of magnetic field on this transition.

\textit{3.2.2.2 Specific heat}

To further check the nature of the transition at $\sim$240K we measured the specific heat around it on sample $\textbf{A3}$. If the observed anomaly  were due to a structural transition having macroscopic effects on the electrical resistance and the magnetisation, we should expect a visible anomaly at $\sim$240K, as with, e.g. a spin or charge density wave(CDW)\cite{Harper}. Within experimental error there is no anomaly as shown on Fig.~\ref{HyCy240}f, which excludes the hypothesis of an structural transition.

\textit{3.2.2.3. de Haas-van Alphen measurements}

Analysis of the de Haas/van Alphen oscillations was performed on sample $\textbf{A4}$. They show two distinctive frequencies of 70 and 22 T . These frequencies are completely different than the ones of Bernal graphite\cite{Piot}, indicating much bigger Fermi surfaces of 68 and 21 x 1012 cm$^{-2}$. By analysing the temperature dependence of these oscillations, an estimation of the effective mass for these carriers can be obtained giving 0.065 and 0.045 $m_e $, respectively (Appendix C). Most probably they correspond to the bulk average phase and not to the particular superconducting grains.

\textit{\textbf{3.2.3. 320K Transition.}}  

We show the ZFC-FC measurements on sample $\textbf{C1}$ on Fig.~\ref{ZYB}c. The moment has a ferromagnetic  behaviour at high temperatures, that we fit with  expression (\ref{Ferro}), giving T$_0$=450K and  a $M_0$ corresponding to 1 $\mu_B$ per 4x10$^{5}$ carbon atoms. When we subtract this ferromagnetic dependence from the raw data we obtain the curves shown on Fig.~\ref{ZYB}d. We observe a transition to a diamagnetic state at T$_c$=320K, with no other transition. 

On Fig~\ref{ZYB}e we show the magnetisation as a function of field measured at 5K after cooling from 400K. We observe a linear diamagnetic decrease followed by a paramagnetic increase, with a minimum at about 0.2T.

We performed magnetic moment hysteresis cycles at different temperatures. We observe cycles (not shown) similar to those of Fig.~\ref{R-M}d.  Subtracting the observed linear paramagnetic behaviour as with sample $\textbf{A1}$, diamagnetic cycles are apparent. They are shown on 
Fig.~\ref{ZYB}f. $\Delta$M(H,T) is extracted from each cycle, and plotted as a function of temperature. The resulting curve, plotted on inset of Fig~\ref{ZYB}f, yields T$_c\sim$320K, as obtained from the ZFC-FC curves.

\textbf{4. DISCUSSION}

As graphite has the strongest known diamagnetic response, the hypothesis of structural phase transitions rearranging the graphene planes, thus recovering a strong diamagnetism (that at 5K corresponds to $\sim$1$\%$ of that of graphite) must be considered. However, the huge resistivity jump cannot be easily explained with this hypothesis, nor the diamagnetic hysteresis cycles or ferromagnetism. Also, structural transitions yielding such strong effects should be visible in the specific heat. While small amounts of superconductivity will not be measurable in the specific heat, which is a bulk measurement, in spite of being able to shortcut the electrical resistance through a tiny filamentary array.

Small resistivity drops  at similar temperatures have been also observed in graphite intercalation compounds with K, Rb and Cs\cite{Fischer}, that were associated with intra- and interlayer alkaline metal ordering effects, although not all the order transitions coincided with the drops\cite{Fischer}. Magnetic susceptibility measurements showed small anomalies with very different shapes at the transitions, 
differences that hindered a full interpretation of the origins of these jumps\cite{Mareche}, that were not studied further. Recent measurements\cite{Mukai} on Li$_x$C$_6$ even show small unexplained magnetisation drops at 300K, together with paramagnetic to diamagnetic changes as a function of \textit{x} and ferromagnetic cycles at certain \textit{x}.

On the other hand, strong evidences of unidentified microscopic amounts of superconductivity in stacking faults of different types of graphite have been reported \cite{Esquinazi}. These claims undeniably show very small amounts of  SC whose precise determination by, e.g. Meissner effect, is prevented by the large diamagnetism of graphite. 

Thus, as high temperature superconductivity might exist in graphite, we are lead to hypothesise that the observed diamagnetic transitions are due to superconductivity. 

In strong support for this hypothesis we have the behaviour shown on Fig.~\ref{R-M}c and Fig.~\ref{ZYB}e, that can only be easily explained by superconductivity, as at low temperatures the ferromagnetism with T$_0$=450K is already saturated, and a graphite diamagnetism alone is not expected to give a minimum. Also, the magnetisation cycles that we show on Fig.~\ref{R-M}e, Fig.~\ref{HyCy240}d and Fig.~\ref{ZYB}f are specific of superconductors.

We can now go further in our analysis and try to elucidate the origin of superconductivity. Unfortunately, for the time being, we are restricted to speculative grounds.

According to our crystallographic analysis our samples are formed by small crystallites($\sim$100nm) that scan stackings of six-fold (or five-fold) slabs with all possible angles (Fig.~\ref{Crys}d). We can hypothesise that a very small number of these crystallites will have the appropriate structure for being superconductors ("magic angle"). Superconductivity can then be attained by the suitable doping yielded by the locally variable potassium content of our samples.  

According to this hypothesis, the superconductivity that we observe is of granular nature and located in twisted potassium occupied interfaces of the type of Fig.~\ref{Crys}d. The small superconducting regions are then grains of platelet type with a ratio diameter to thickness $\geqslant$1000. Besides, as our crystallites are in the 100nm range, we expect a large field penetration. In this case, the standard estimation of the superconducting volume obtained by comparing the negative slope of the low field region of the magnetisation at low temperatures of a ZFC sample to perfect diamagnetism will largely undervalue it. As we lack another way to estimate the superconducting volume, though, we nevertheless do this comparison with the slopes of Fig.~\ref{R-M}c and Fig.~\ref{ZYB}e. We obtain a superconducting volume of $\sim$ 10$^{-5}$ and 5x10$^{-6}$ of the total volume, respectively.

Our samples scan a great number of stacking possibilities, but very few will have the conditions to be superconducting. We try here to evaluate what is the expected superconducting sample volume assuming the "magic angle" hypothesis. By symmetry, the possible inequivalent angles range\cite{Mayou} from 0$^\circ$ to 30$^\circ$. The probability of having a crystallite with the "magic angle" is thus 1/30 (considering that an 1$^\circ$ error is tolerable). Statistically then,  the "magic angle" crystallites will be separated in each direction by 30 non-superconducting ones. So, if all the angles have the same probability, we expect that only (1/30)$^3$$\sim$4x10$^{-5}$ of the volume will be occupied by "magic angle" crystallites, that will be superconducting only if they have the appropriate doping. This value is of the same order of magnitude of the previous estimation.  The fact that we do observe measurable superconducting-like diamagnetic cycles and even almost obtain percolation to zero resistance, means that the probability of having small "magic angles" is greater than the 1/30 that we have hypothesised and that estimated superconducting volume is highly undervalued.

Although it remains to be demonstrated, the ferromagnetism that we observe may also be intrinsic, as the ferromagnetism of HOPG samples similar to those from which we develop our synthesis has been studied and found not to be due to ferromagnetic impurities\cite{EsquiFerro1,EsquiFerro2}. Thus, a certain number of these crystallites may have a doping suitable for a ferromagnetic ground state, which may explained the observed ferromagnetism.

The crystallographic disorder of our samples favours twists defects at the AA interfaces with low potassium occupation, presumably rendering six or five graphene AB slabs with twist defects. We may speculate that our different samples show the fixed values of T$_c$'s $\sim$110K, $\sim$240K and $\sim$320K that would indicate the lowest "magic" angles for six-fold twisted slabs. This case has  been treated theoretically for two infinite slabs, yielding very flat bands\cite{Guinea}. While the number of low "magic angles" for n layers increases as n/2, with three expected for six-fold slab stacking\cite{Hierarchy}.

\textbf{5. CONCLUSION}

We have developed an intercalation-deintercalation (IDI) process of graphite that yields a highly disordered six-fold slab graphene stacking, with arbitrary angles and inhomogeneous potassium atom distribution. The disordered doping and twisted stacking nature of the structure of our samples, by probing all type of defects including twists, may be at the origin of the different diamagnetic transitions and possibly also of the ferromagnetic phases.  Besides we should note that the most immediate and simple explanation that englobes all the observed data is superconductivity. Although the observed superconductivity is small, its existence is systematic in many samples. Besides, the small measured value is what is expected if we have an statistical distribution of crystallites with different Moiré angles ranging from 1$^\circ$ to 30$^\circ$, with only the smallest angle yielding superconductivity. Furthermore, the IDI process unveils these transitions by annihilating graphite diamagnetism through the breaking down of the stacking of AB graphene planes to six or five fold slabs, together with the doping by potassium. Our results add on previous reports of room-temperature superconductivity in graphite\cite{Esquinazi}. Further work is necessary to improve the process (different alkaline element, temperature of deintercalation, etc.), in order to demonstrate the superconducting nature of the diamagnetic transitions by obtaining zero resistance and controlling the T$_c$. As in our first efforts we obtain an almost percolating sample for the $\sim$240K transition and only one transition for the $\sim$320K sample, this objective seems attainable.

 \textbf{Acknowledgments}

MNR thanks  K. Hasselbach, J-L. Tholence, C. Paulsen, D. Mayou, A. Aligia for discussions and  J-L. Tholence, P. Monceau, J.E. Lorenzo-D\'iaz and K. Hasselbach for critical reading of the manuscript. We acknowledge support from the French National Research Agency through the projects IRONMAN ANR-18-CE30-0018 and PRESTO ANR-19-CE09-0027.

\textbf{Appendix A: \textit{Crystallography}}

\begin{figure}[H]
\includegraphics[width=\linewidth]{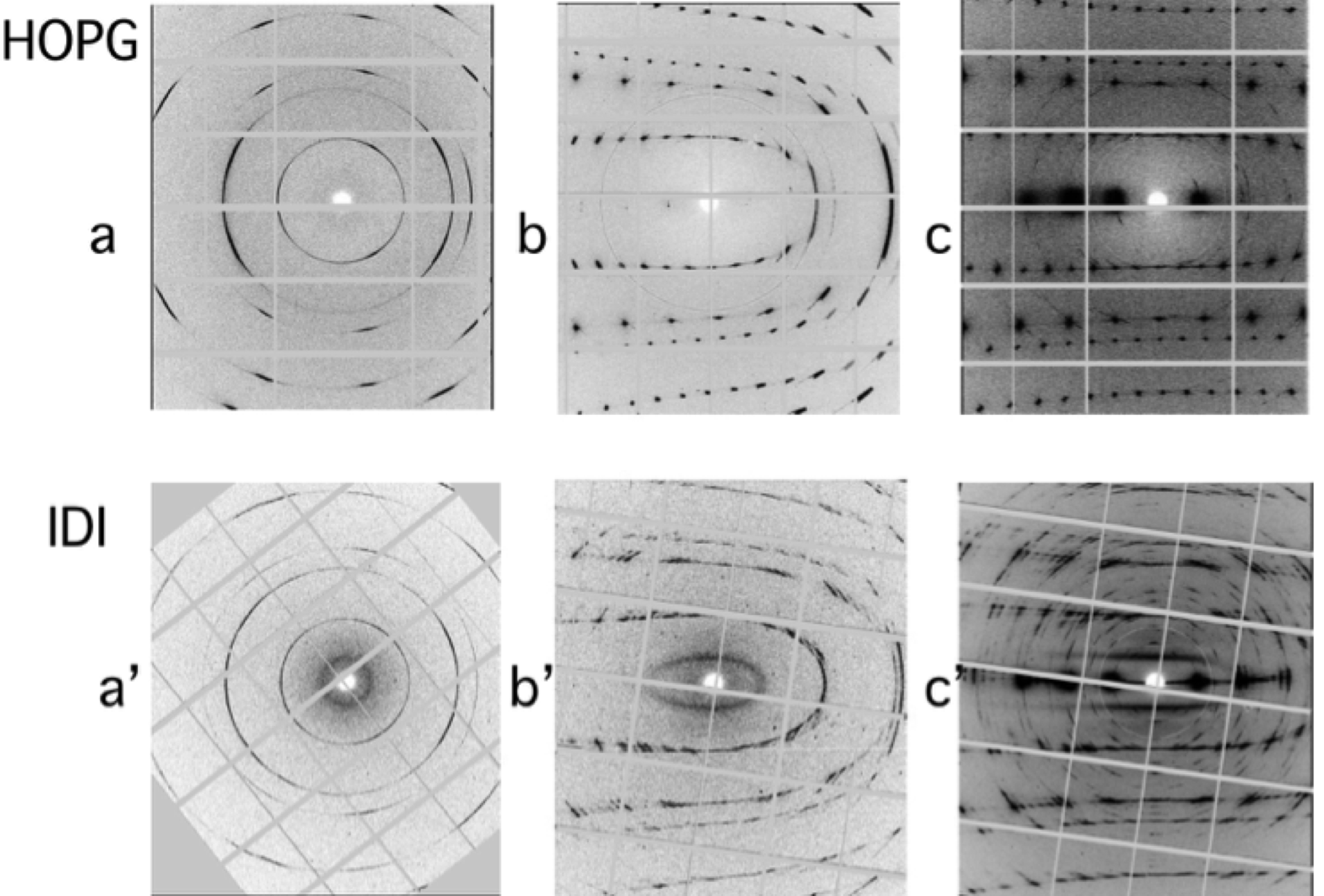}
\caption{ Spectra from an untreated HOPG sample (a, b, c) and of IDI new carbon allotrope (a’, b’, c’), which evidence common scattering-diffraction intensities corresponding to a same basic sub-structure: 
(a) and (a’) patterns corresponding to orientations close to ($\textbf{ab}$) planes ($\omega$ about 0-10$^\circ$) they present, in both HOPG and IDI phases, the same full rotation of ($\textbf{ab}$) planes due to disorder of their stacking along $\textbf{c}$ axis, they give  $\textbf{a}$=$\textbf{b}$=2.46 \AA parameters ; 
- (c) and (c') patterns correspond to orientations nearly parallel to $\textbf{c}$ axis ($\omega$ about 80-90$^\circ$) in both HOPG and IDI phases, they present the same intensity distribution of strong Bragg reflections, but the planes’ stacking is different, while in the HOPG phase it corresponds to an AB stacking with a $\textbf{c}$= 6.70 \AA parameter,  in IDI carbon allotrope it presents a modulation along $\textbf{c}$ whose parameter is mainly 22.15\AA(and also 19.0\AA and some disordered stacking in some parts of the IDI sample).
- (b) and (b') patterns of inclined platelet ($\omega$ , about 55-60$^\circ$) which evidence an analogue disordered stacking in the HOPG sample and in the new carbon allotrope . 
Additionally in (a’,b’,c’) patterns, the IDI presents additional diffuse scattering which corresponds to a short range order ($\sim$6 \AA) (SRO) in the ($\textbf{ab}$) planes and a total disorder of its stacking along $\textbf{c}$ : this SRO gives a diffuse ring in Fig a’, two broad diffuse lines parallel to the $\textbf{c}$ direction in Fig 1c’ and a deformed “ellipse” in Fig. b’ which correspond to the intersections of a section of a diffuse cylinder corresponding to a $\sim$6 \AA~SRO in the ($\textbf{ab}$) planes and a total disorder along $\textbf{c}$ direction.
}
\label{Crys1} 
\end{figure}

Highly oriented pyrolytic  graphite is a textured material made of platelet-like crystals whose $\textbf{c}$ axis are parallel within  about 1$^\circ$, but with a disorder considering $\textbf{a}$ and $\textbf{b}$ axis. This characteristiccan be observed in X-ray diffraction patterns by the occurrence of scattering circles in the (a b) planes. To analyse the relationship of our new IDI modulated phases with the original HOPG one, we have performed our diffraction experiments on both of them. For X-ray diffraction we have analysed one of the HOPG sample used as starting material to the K Insertion Des-Insertion process. As seen in both Fig.~5abc, a’b’c’ and Fig.~6 a, bb’), there is a close relation in the distribution of diffraction intensities between the starting HOPG material and the obtained K-IDI phases which are partly disordered but present 
some characteristics of the ordered Graphite Inserted Compounds such as the KC$_y$ allotropes \cite{Schulze}    but in our case with a longer modulation KC$_{72}$  or KC$_{60}$.

We observed on Fig.~5 patterns, the same continuous scattered circles in $\textbf{ab}$ planes, which characterise the presence of the disordered stacking of graphite planes in both HOPG and transformed IDI allotropes. A stacking modulation along $\textbf{c}$ axis appears clearly on Fig.~5b’c’ and on Fig.~6bb’ showing a similar stacking type of carbon planes (with different $\textbf{c}$ values) in HOPG and for the treated IDI sample. Additionally, the IDI presents additional diffuse scattering which corresponds to a short range order ($\sim$6 \AA) (SRO) in the $\textbf{ab}$ planes and a total disorder of its stacking along $\textbf{c}$ (Fig.~5a’,b’,c’). This SRO, due to Potassium atoms, as proposed in some ordered Graphite Inserted Compounds\cite{DiCenzo,Clarke,Moret,Mori} for stages L$\geq$2, gives rise to the diffuse ring in Fig.~5a’, the two broad diffuse lines parallel to the $\textbf{c}$ direction in Fig.~5c’) and the deformed “ellipse” in Fig.~5b’). It corresponds to a $\sim$6\AA ~ SRO in the $\textbf{ab}$ planes and a total disorder along $\textbf{c}$ direction.
Due to its size, the K atoms are in between two carbon hexagons (above A and below A’). 
The distance in between these two A and A’ carbon planes is large (about 5.2-5.6 \AA), favouring rotation disorder. 
\begin{figure}[H]
\includegraphics[width=\linewidth]{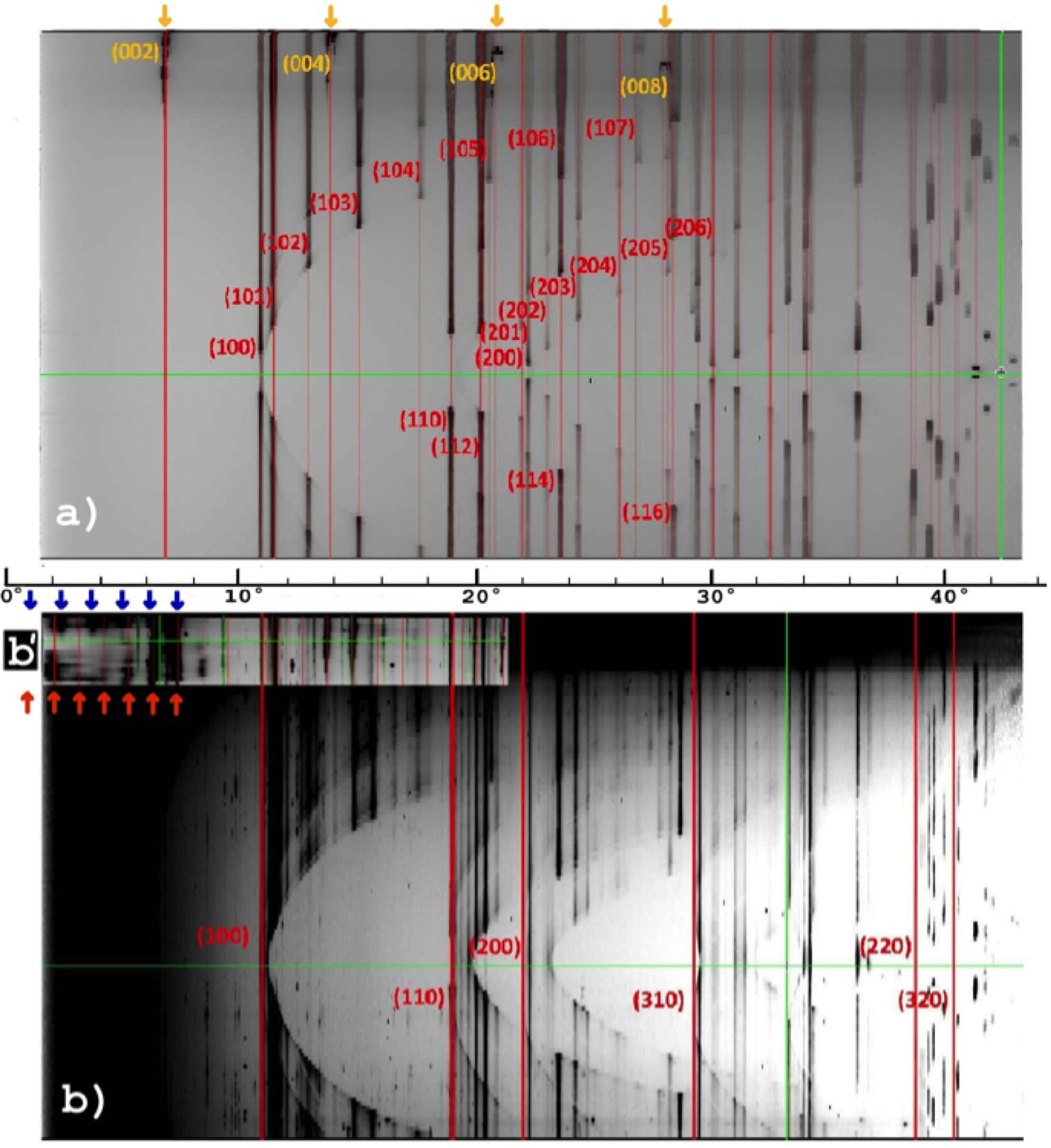}
\caption{  Sum of all diffraction-diffusion ($\omega$, 2$\theta$) patterns (with $\omega$ from 0$^\circ$ to 90$^\circ$) of the untreated HOPG sample (a) and of the IDI allotrope (b,b’) : they show common intense Bragg spots, corresponding to a same basic sub-structure, suggesting  ordered models for a K ABABAB K BABABA (or K ABABA K ABABA) stacking and for a SRO distribution of K atoms in ($\textbf{ab}$) planes associated with a disordered distribution of K atoms along the $\textbf{c}$ stacking direction :
- (a) and (a’) ($\omega$, 2$\theta$)  patterns corresponding the sum of all patterns ($\omega$ from 0$^\circ$ to 90$^\circ$) which evidence the 2 x 3.35\AA stacking of HOPG sample (yellow small arrows) and (b,b') of the 2 x 22.15 \AA (but depending to the analysed sample area, it can be locally 19.0 \AA or partially disordered) stacking of the carbon IDI allotrope (red and blue small arrows). This analogue intensity distribution of diffraction/diffusion in both HOPG  and  IDI allotrope implies a same AB stacking for both phases;  the 2$\theta$ positions of ($hkl$) scattered intensities at medium and long angles in the modulated allotrope agree with a long range order of the structural distortion along $\textbf{c}$. This intensity distribution can be related to a K ABABAB K BABABA (or K ABABA K ABABA) stacking in the  IDI carbon allotrope (to better focus on the possible local variation of the $\textbf{c}$ parameter of the modulation , we have plotted the (b’) part of the pattern with a low saturation allowing to evidence the 6-fold modulation 2x 22.15\AA (red arrows) together with the 5-fold modulation 19.0\AA (blue arrows), the SRO concerns mainly the distribution of K atoms in between two AA graphene planes,  contributes at low 2$\theta$ angle, in the dark part of the Fig.~A2b.}
\label{Crys2} 
\end{figure}
The SRO diameter of the diffuse cylinder, observed on IDI allotropes, corresponds to a value $\sim$6 \AA 
which indicates that it probably originates from the K-K liquid-glass distribution\cite{DiCenzo,Clarke,Moret,Mori}. This K insertion favours rotations of AA’ carbon planes ( Fig.~7 and  Fig.~1e). Since
this stacking must be done with a lower distortion, as the inserted K atoms favour the superposition of “low” hexagons A above “up” hexagons A’ .
As a consequence,  parallel or rotated orientations of carbon planes A and A’ are favoured  (black arrows in Fig.~7). 
These also induce  twisting errors of these stacked graphite planes\cite{Guinea}  . 
Such errors in rotations of graphite planes would also cause the occurrence of some local distortions in the planes.

\begin{figure}[H]
\includegraphics[width=\linewidth]{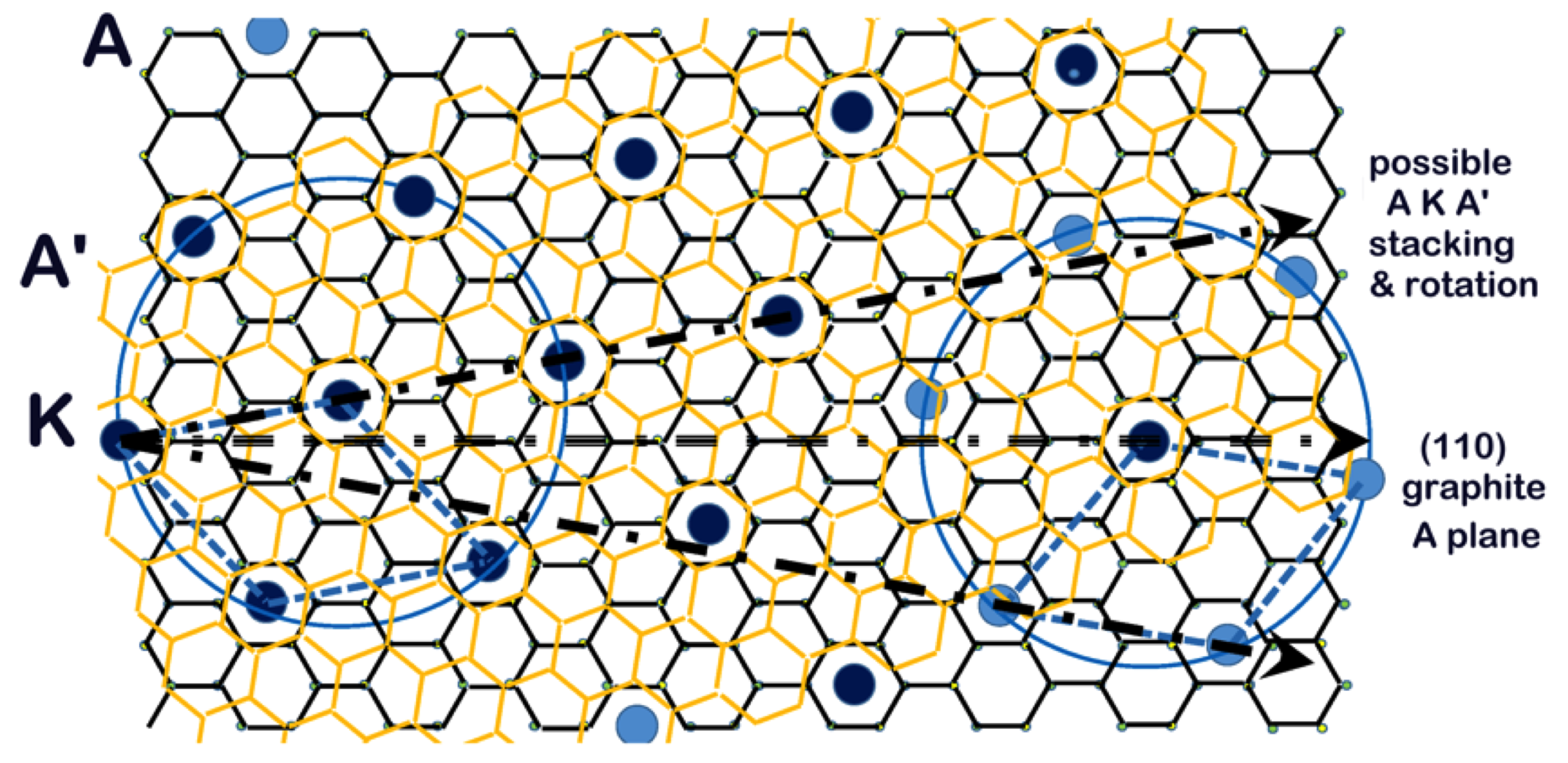}
\caption{ Proposed example of stacking of graphite AA’ planes at the interface of IDI carbon allotropes, where K atoms are inserted (Fig 1de). Due to its size, the K atoms are most probably in between two carbon hexagons (above and below) ; such a stacking must be done with a lower distortion which favours some orientations of carbon planes A and A’, either parallel or rotated with orientations which induces the superposition of “low” hexagons A above “up” hexagons A’ (arrows in the figure). Due to potassium atom size the distance in between these two A and A’ carbon planes must be quite large (about 5.2-5.6\AA) and favours these possible rotations but also some few degrees twisted errors of these carbons’ planes stacking. Furthermore the observed diffuse central cylinder (Fig 1a,b,c) supports a SRO of  potassium distribution in the ($\textbf{ab}$) plane, in- between AA planes, and a total disorder along the $\textbf{c}$ direction and also a disordered rotation in the ($\textbf{ab}$) plane ; this SRO also could induce some few degrees twisted errors of these carbons’ plane stacking.}
\label{Crys3} 
\end{figure}

\newpage 

\textbf{Appendix B: \textit{Raman}}

\begin{figure}[H]
\centering
\includegraphics[width=\linewidth]{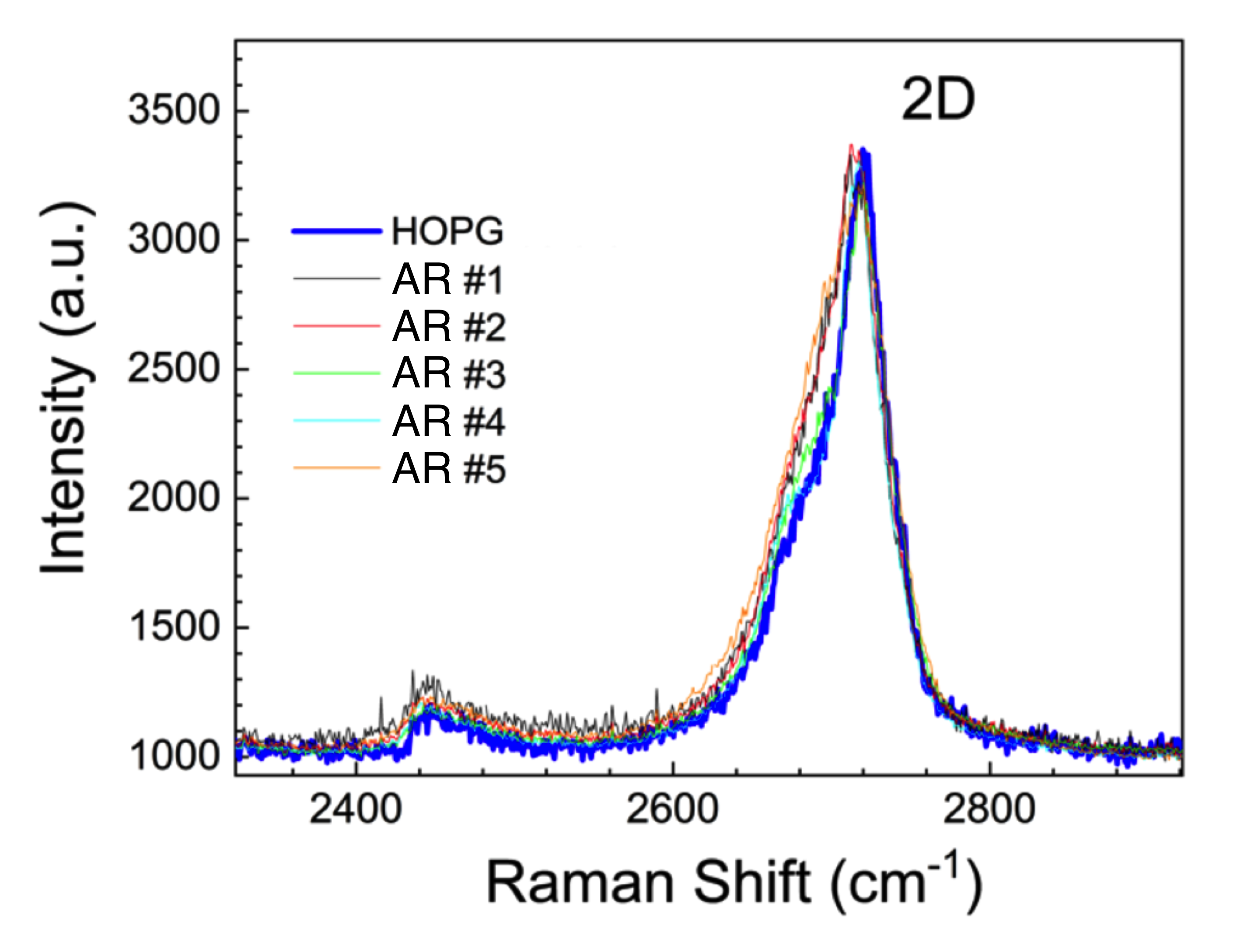}
\caption{  Raman spectra of the 2D band of HOPG graphite before (blue) and after intercalation/desintercalation proccess sample \textbf{AR} measured on different spots. The spectra are normalised to the maximum of the 2D mode. A spatial-dependent slight broadening of the line is measured after the process.}
\label{Raman2D} 
\end{figure}

\textbf{Appendix C:\textit{ Effective mass estimation from de Haas-van Alphen}}

The quantum oscillations can be expressed by just the first harmonics in the moderate magnetic field limit (3).
\begin{equation}
dM\propto D_Texp \left (-\frac{\pi}{\omega_c\tau_e} \right )cos \left(\frac{2\pi F_B}{B}-\varphi \right )
\end{equation}
Where $\omega_c$=(e B)/m$^*$  is the cyclotron frequency, m$^*$ is the cyclotron mass,$F$ is the magnetic frequency of the oscillations, and the prefactor D$_T$=$\gamma$/(sinh⁡($\gamma$)) with $\gamma$ = (4$\pi^2$ k$_B$ T)/ $\hbar$  $\omega_c$ ) takes into account the amplitude temperature dependence of the oscillations.

By performing field sweeps at constant temperature we can easily determine the oscillation frequency as shown on Fig.~9.

\begin{figure}[H]

\includegraphics[width=\linewidth]{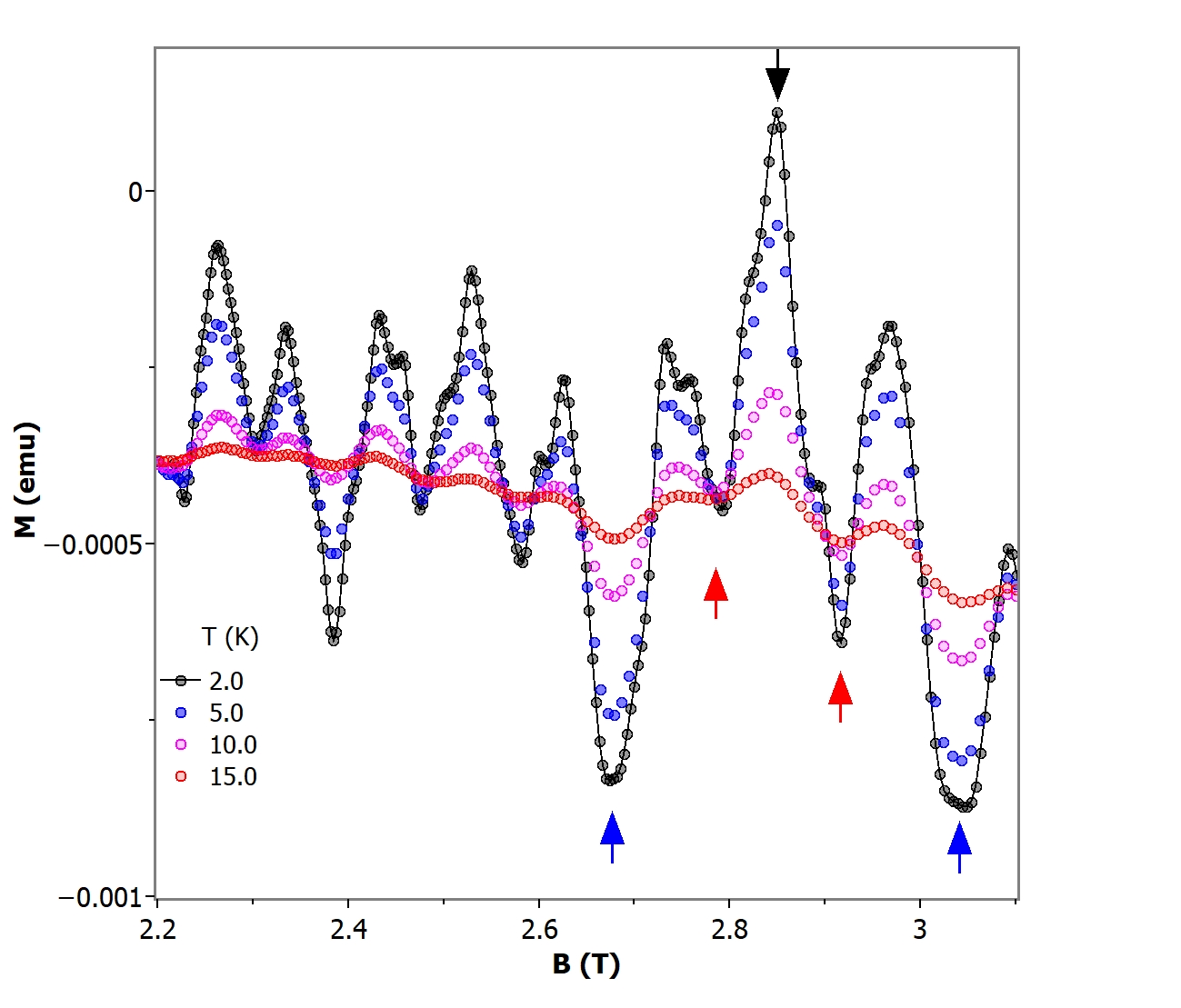}
\caption{The arrows indicates the selected fields at which magnetization as function of temperature has been measured. By considering the fields indicated by Black and Red arrows only the high frequency amplitude dependence will be obtained. For the low frequency temperature dependence, the fields indicated by Black and Blue arrows should be considered and substracting the previously obtained high frequency amplitude.}
\label{Haas1} 
\end{figure}

Conversely, to estimate the effective mass, we should do temperature sweeps at constant field. In this way the exponential prefactor and the oscillatory part are constant, addressing directly the prefactor $D_T$. 
Because two magnetic frequencies are present in this system (called H for High and L for Low), we have carefully chosen the magnetic fields values as to don’t be affected by the modulation of the another one.
It can be seen in the figure Fig.~9, that at some values of magnetic field the two frequencies are in phase (Black and Blue arrows) and for others out of phase (Red arrows). We can then state that the amplitudes are:

Therefore:
\begin{equation}
A_{Black}=A_H+A_L
\end{equation}
\begin{equation}
A_{Red}=-A_H+A_L
\end{equation}
\begin{equation}
A_{Blue}=-A_H+A_L
\end{equation}
\begin{equation} 
A_H=\frac{A_{Black}  - A_{Red}}{2}
\end{equation}
\begin{equation} 
A_L=\frac{A_{Black}  - A_{Bleu}}{2}-A_H	
\end{equation} 		
The obtained amplitude is then normalised by its lowest temperature value and plotted in  Fig.~10. A fit by expression $D_T (B,T)/D_T (B,0)$=$\frac{\gamma}{sinh⁡(\gamma)} $gives us the only fitting parameter: the cyclotron mass $m^*$.

\begin{figure}[H]
\centering
\includegraphics[width=\linewidth]{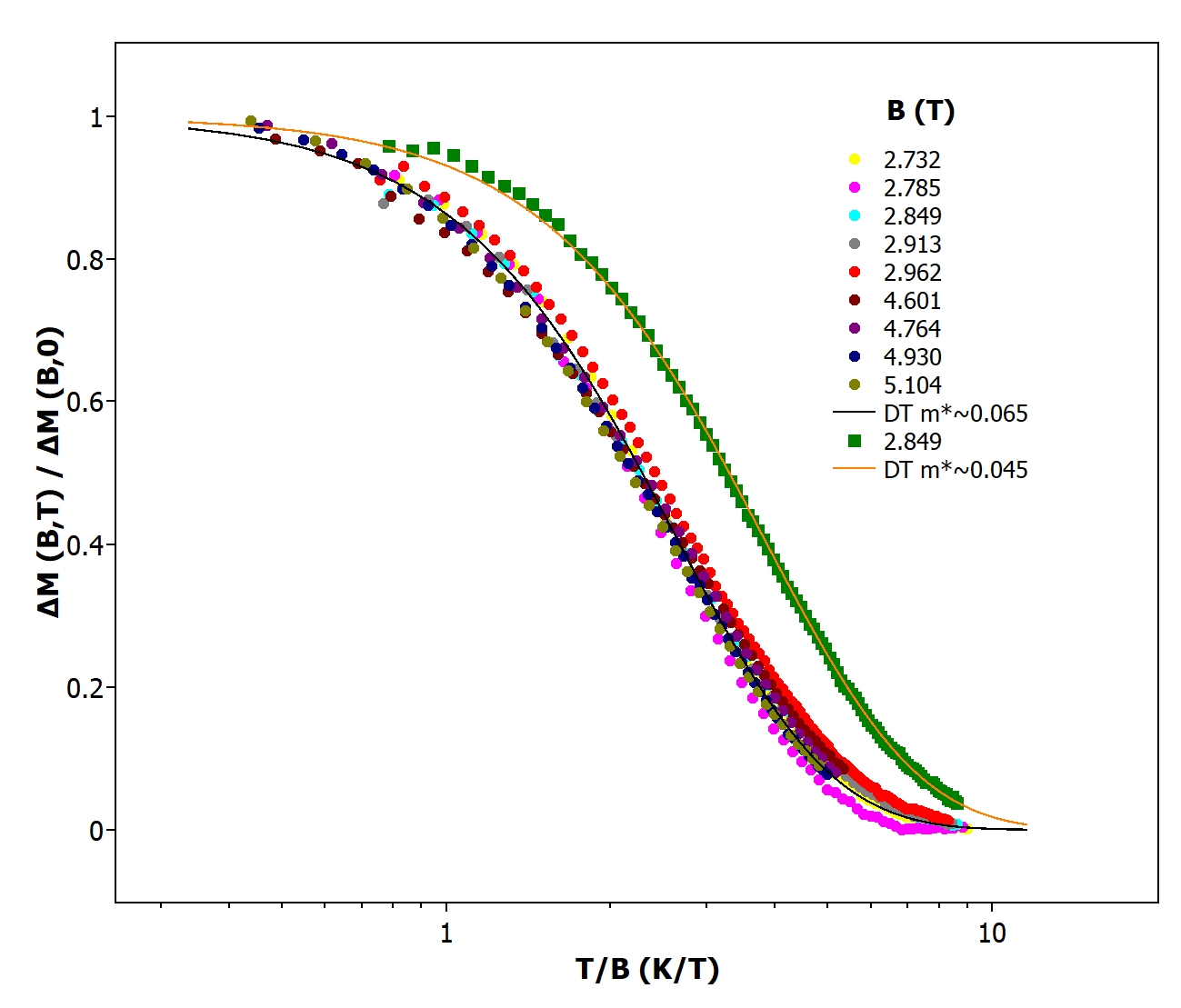}
\caption{  Temperature dependence of each magnetic frequency amplitude for constant fields. The expressions 2 had been used to obtain only one magnetic frequency amplitude. While many values of field (from 2.732 till 5.104 T) had been used for the higher frequency, only one (2.849 T) was used for the lowest frequency.}
\label{Haas2} 
\end{figure}

\end{document}